# A Unified E2E Energy Efficiency Testing Framework for Open RAN

Marcin Hoffmann[1,2], Marcin Dryjański[1], Adrian Kliks[1,2], Andreas Gladisch[3],
Ajesh Pulyaar Keerthi[3], Mohammadreza Razmi[3], Heiko Lehmann[4]

[1]Rimedo Labs, Poland; [2]Poznan University of Technology, Poland, [3]i14y, Germany, [4]T-Labs, Germany

**Abstract**

Energy efficiency (EE) is one of the key challenges for contemporary and future mobile networks, including within the Open Radio Access Network (O-RAN) architecture. However, there is a significant gap in common procedures for comparing the EE of both hardware (HW) and software (SW) solutions offered by various vendors. Usually, EE improvements of both SW and HW solutions are demonstrated in a specific scenario defined by individual vendors avoiding comparisons and benchmarking under various network conditions. This paper outlines the need for unified end-to-end (E2E) EE testing for O-RAN. First, it analyzes the standards to identify missing parts. Based on the analysis, a novel O-RAN E2E EE Testing framework is proposed. The framework aims to test the EE of the xApp/rApp pair cooperating on the cell on/off switching using a commercial RAN emulator and real-world network topology data from a mobile network operator (MNO). The test results show up to 57% improvement in EE compared to the baseline.

## I. Introduction

Optimizing energy or power consumption (EC/PC) has been one of the key challenges in wireless communications for many years. The continuously increasing electricity prices and growing environmental awareness have further emphasized the need to improve energy efficiency (EE) in mobile networks. Thus, it also became one of the most important topics within the O-RAN space, where the disaggregation and multivendor aspects add complexity to the network's performance evaluation. It is widely expected that the deployment of proper rApps and xApps on a RAN Intelligent Controller (RIC) should lead to significant energy savings (ES) [1]. Such expectations have been confirmed by numerous setups, e.g., showcased during the O-RAN Global Plugfests, where various solutions have been implemented and tested [2].

However, the conducted experiments have focused on improving EE in specific situations and aspects. In detail, hardware (HW) vendors and software (SW) providers typically demonstrate EE gains of their products using their own evaluation setups and scenarios. Thus, the obtained EE gains may vary if, e.g., the traffic profile and network topology are modified. Also, the target deployment platform, along with its computation performance and delays between various network modules, can affect the resultant EE. Moreover, usually, the HW and SW vendors compare their solutions against baseline algorithms, avoiding direct and comprehensive benchmarking of their solutions. In addition, the considered setups are typically limited to a single component stack for HW evaluation, and simulations covering only a few cells for SW solutions. Thus, those tests do not consider the broader impact of the solution on other parts of the network. Finally, contemporary networks progress to virtualization. Therefore, the realization of functionalities in a virtual environment needs to be taken into account when drawing conclusions on the network's EE. More specifically, conducting EC/PC measurements of a mobile network composed of physical and virtual network functions (PNFs and VNFs), along with the possibility of deploying VNFs at various cloud locations, becomes a challenge.

The above discussion led to the need for a unified end-to-end (E2E) energy testing framework, which can be used as a benchmarking environment for EE testing of O-RAN network components and applications. Comparing individual O-RAN elements (HW or SW entities) can be done with smaller setups, e.g., a few User Equipments (UEs), and a few Base Stations (BSs). However, for comparing or measuring the performance of x/rApps related to ES, scale should be involved, resulting in the need for larger scenarios. Going further, to capture the cloudification aspects (like deployment options), the framework should also encapsulate different methods for infrastructure EC/PC measurements (covering cloud, VNF, and PNF aspects). In addition to the benchmark setups, there should be widely accepted procedures for: verifying the actual network elements EC/PC; testing the EE improvement achieved by an rApp or xApp; or measuring the overall energy usage in the network. To date, multiple standardization bodies have contributed to this topic. However, their activities are not aligned with each other, e.g.:

- ETSI proposes methods to evaluate the EC of a BS [3]-[5], but lacks the O-RAN context.
- 3GPP does not define the E2E EE evaluation

procedures but has a lot of reference scenarios and mathematical models, which could be useful during large-scale simulations [6]-[9].
- O-RAN ALLIANCE aims to provide a test framework for EE use cases [10]-[13] while lacking large-scale scenarios for x/rApps evaluation under unified conditions.

Considering the results of the analysis of the building blocks covered by standardization bodies and the identified gaps, the main aim of this paper is to propose a unified O-RAN E2E EE Testing Framework. It aims to serve as a platform to enable comparing both HW and SW solutions and their impact on the EE of an O-RAN network. The paper is organized as follows. First, it provides a review of EE-related aspects from the perspective of the documents oriented on EE measurement and evaluation from organizations like ETSI, 3GPP, O-RAN ALLIANCE, and NGMN. Based on those inputs and identified gaps, the paper proposes a high-level E2E EE Testing Framework discussing its features and applications in various network scenarios and use cases, considering different aspects of EE testing. This is followed by the example realization of the proposed E2E EE Testing Framework to test xApp and rApp under a large-scale network emulated with the use of real-world data provided by an MNO. The conclusions provide recommendations for the envisioned future directions.

## II. Current Status of Standardization Efforts

The EE in mobile networks is a broad topic covered by multiple organizations and standardization bodies. Unfortunately, distinct aspects of EE testing and measurement are covered by different ones. Throughout this section, we provide an overview of the standardization effort toward the E2E EE testing framework, to identify the maturity of the work and missing parts from the perspective of ETSI, 3GPP, NGMN, and O-RAN ALLIANCE.

### *A. ETSI*

ETSI covers a significant part of the work already done, and other organizations often refer to it. Based on the content of ETSI specifications related to testing and measurement of the BS EC/PC, the following conclusions were formulated:

- From the O-RAN perspective, the BS EC/PC models used in [3]-[5] should be adjusted to the 7.2 split, resulting in O-CU, O-DU, and O-RU entities in various configurations, including options with virtualized O-CU/O-DU and RIC, not considered by ETSI documents to date.
- For both static and dynamic power measurement (see [3] and [4] respectively), an MNO is allowed to define a load distribution profile reflecting the situation in the network and mandate that profile to be used. From the O-RAN perspective, it might be beneficial to define several load profiles reflecting scenarios, e.g., urban, suburban, and vehicular. The best would be to utilize MNO real-world data.
- The measurement report templates are well-defined and could serve as a strong basis to be extended to cover E2E energy testing in the context of O-RAN. However, here only one BS is tested while there is a need for testing a bigger part of the network to capture a broader view of its EE. Also, the KPIs should be further discussed to take QoS aspects into account.

To sum up, ETSI specifications provide well-defined procedures for the measurement of BS PC, including, e.g., temperature, voltage requirements, UE traffic models, and distribution. However, they lack O-RAN architecture context, procedures for cloud deployment of NFs, and scenarios for evaluation of the large-scale networks, necessary, e.g., for evaluation of x/rApps controlling tens of cells.

### *B. 3GPP*

In contrast to ETSI, 3GPP specifications do not contain the exact procedures for EE testing. They cover scenario definitions, which could be utilized for x/rApp evaluation. Based on the EC/PC-focused 3GPP specifications and technical reports (TS/TR) [6]-[9], the following conclusions can be formulated:

- The models proposed by 3GPP documents (including, e.g., the EC/PC model for UE and BS, radio channel models, UE mobility models, and reference network deployment scenarios) can be highly usable while testing the EE gains provided by x/rApps under a large network emulated by a dedicated SW.
- 3GPP indicates a need to enhance the EE KPIs to be more comprehensive in different network scenarios. It is needed to consider other performance dimensions than only the data volume and EC/PC that might blur the QoS of individual users.
- In the context of the E2E EE testing framework, several scenarios with various UE services should be considered, e.g., file download or streaming (i.e., EC/PC measurements should be related to the media consumed by the UE).
- 3GPP EE KPIs and metrics can serve as a potential measurement option for VNF EE.
- The 3GPP proposed EC/PC model, providing a relation between the consumed power and the number of active antennas, spectral density, and utilized bandwidth, can be extended through the measurement of the O-RAN stack.
- Based on 3GPP for the system-level simulations, the network's EE should be evaluated

under at least 2 deployment scenarios: one coverage-limited environment (e.g., Rural) and one capacity-limited environment (e.g., Urban).

Summarizing, 3GPP technical reports provide models that can be used for large-scale, system-level simulations to test xApps and rApps under unified conditions, e.g., PC models, and reference deployments (like urban, rural, V2X). However, there are no procedures for E2E EE/EC testing, as well as O-RAN aspects missing.

*C. NGMN*

While ETSI provided background for HW testing and measurement, and 3GPP documents are related mostly to simulation and modelling aspects, NGMN touched upon virtualization and VNF-related power measurements. Based on the analyzed document [14], the following was observed:

- O-RAN opens up the option of standard commercial off-the-shelf (COTS) HW hosting parts of the RAN functionality, implemented as VNFs, while others as PNFs. This architectural solution is not yet fully covered by metering standards.
- NGMN addresses metering aspects in a virtualized RAN (vRAN) infrastructure, e.g., metering of performance and EC/PC at the HW level, options to determine the EC/PC of a VNF, and impact of cloud operation models on the capability to optimize performance and EE of vRANs. However, it is missing the O-RAN context, like O-RAN functions to be hosted in the form of VNFs, as well as standardized deployment options.
- The discussed open-source solutions for determining EC/PC of VNFs seem promising. However, the accuracy strongly depends on the availability of reliable measurements at the HW level and on the modeling, algorithms used by those solutions. Therefore, without standardization or unification of those models and tool configuration, creating benchmarks would likely not be possible, subject to configuration mismatches, flexibility of data model, and individual settings.
- NGMN suggests using specific industry-adopted solutions (like Kepler, Redfish, Prometheus) for HW measurements, which, however, do not fit into a general framework, being very specific.
- The NGMN documents serve as a good input to the framework in telco virtualization and cloudification, but they are not normative specifications, nor do they provide measurement procedures and scenarios.

To sum up, NGMN is focused mostly on the cloud/virtual deployment of mobile networks and PC measurement of VNFs (per Kubernetes cluster, Pod, container). However, NGMN is missing the O-RAN context and does not provide technical specifications.

*D. O-RAN ALLIANCE*

O-RAN ALLIANCE bases its analysis on the O-RAN specifications while often referencing ETSI and 3GPP. It also touches VNF, but those discussions are not mature enough to become a concrete framework capturing all the aspects. Based on the related O-RAN documents [10]-[13], the following conclusions were drawn:

- Concerning EE, O-RAN only covers rApp-based control of the full system. Others (3GPP, ETSI, NGMN) deal either with full RAN or cloud only, missing the individual elements, testing, and benchmarking, covering the O-RAN full stack.
- Only one-use case is fully defined for testing, i.e., Cell Off/On Switching (COOS) using rApp, while the framework should be more general (i.e., capturing all ES-use cases and xApp-controlled options as well).
- The cloud-related ES use case is still a work in progress at the O-RAN level. This topic, from the perspective of measurements, modeling, and procedures, is new and not yet fully defined in O-RAN.
- The performance of the system with RIC optimizations shall be benchmarked through analysis of KPI measurement results. The performance gain with RIC optimization is evaluated by comparing the test result with RIC optimizations enabled versus that with them disabled.
- O-RAN defines functional and performance tests for EE under the COOS use case. The test assumes a typical deployment of O-RU as PNF and O-DU, O-CUs as VNFs in O-Cloud. The performance tests utilize the ETSI traffic model for dynamic power measurement.
- For E2E EE testing of x/rApps in O-RAN, the deployment scenarios defined by 3GPP can be considered.

To summarize, O-RAN ALLIANCE specifications for E2E testing are not fully developed yet. They contain test procedures only for a single use case. The test cases for the large-scale scenarios are still missing – they rely on the ETSI dynamic load measurements that are good for evaluating a single stack composed of O-RU, O-DU, and O-CU. Also, some of the KPIs proposed in the O-RAN ALLIANCE might not be available in practice due to a lack of their implementation in the vendor's solutions. Finally, VNF and cloud aspects are present in O-RAN documents, but they are at an early stage of definition.

### E. Summary of the standardization efforts

After analyzing the documents and specifications from the respective organizations, the main observation is that each entity provides documents containing important notes and (in some cases) normative specifications related directly to the EE framework. However, to produce a comprehensive E2E EE testing framework, the information should be combined, as each of them touches separate, sometimes disjoint aspects.

The scope of the documents defined by each organization, along with their gaps concerning an O-RAN-related EE framework, is summarized in Table 1.

*Table 1 Summary of standardization efforts*

| Org. | Scope | What is missing? |
|---|---|---|
| ETSI | • defines metrics and measurements for BS EE/EC<br>• provides models for BSs and requirements for measurement equipment<br>• describes standardized test setups and procedures for power measurements (static and dynamic) for a single base station<br>• standardized reports from the measurements<br>• good starting point for O-RAN extension | • O-RAN context and architecture, e.g., 7.2 split<br>• scenarios/configurations for large-scale network power measurements<br>• procedures for testing mobile networks with VNF, e.g., virtual CU/DU<br>• QoS in EE definitions |
| 3GPP | • frequently refers to ETSI regarding EE<br>• defines reference scenarios and models for system-level simulations<br>• demand for measurement of VNFs' EE/PC<br>• EE evaluation based on multiple scenarios: MBB, URLLC, mMTC/Rural, urban | • O-RAN context<br>• procedures for E2E EE testing<br>• EE/Power consumption KPIs are not available per HW unit or VNF<br>• most of the material referring to EE/EC is from TRs, not TS, i.e., not normative |
| NGMN | • cloud vs physical deployment for a mobile network<br>• presents aspects relevant to cloud-type/environment measurements<br>• metering servers, storage, and network components<br>• solutions for VNF power measurement<br>• optimization of RAN deployment<br>• points gaps/missing elements between the O-RAN and cloud aspects | • O-RAN aspects<br>• gathers different aspects, but does not fill the gap between the cloud itself and the network<br>• not a specification |
| ORAN | • related directly to O-RAN aspects<br>• defines NES methods, metrics, and requirements associated with the O-RU (COOS and RCR)<br>• concepts, requirements, and use cases optimizing for O-Cloud resources along with metrics (early stage)<br>• test procedure, test equipment definition, test setup and configuration, test criteria for E2E testing for O-RAN ES (COOS) | • details and tech specs for O-Cloud aspects<br>• different maturity stages for different use cases regarding ES and E2E testing<br>• different documents for different parts of the system (treating separately O-Cloud from COOS, RCR, ASM)<br>• test specs only for COOS rApp (early stage)<br>• several options for measurements and KPIs (not all are implemented in practice) |

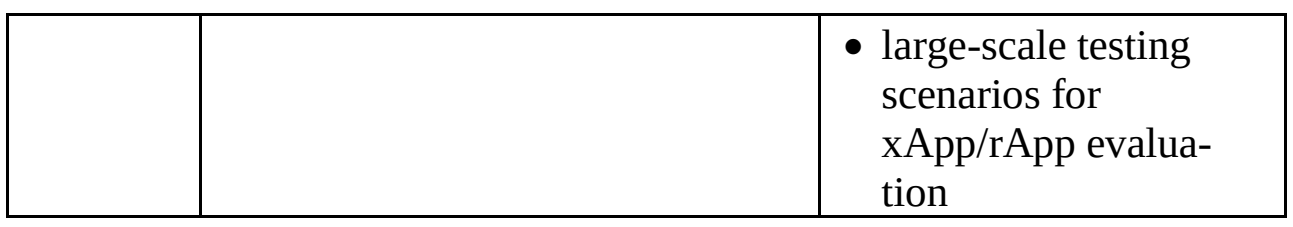

| | | |
|---|---|---|
| | | • large-scale testing scenarios for xApp/rApp evaluation |

## III. The E2E EE Testing Framework

By reviewing the standardization efforts, we identified the key aspects covered by each entity as well as missing parts. Based on these considerations, we propose an E2E EE Testing framework depicted in Fig. 1.

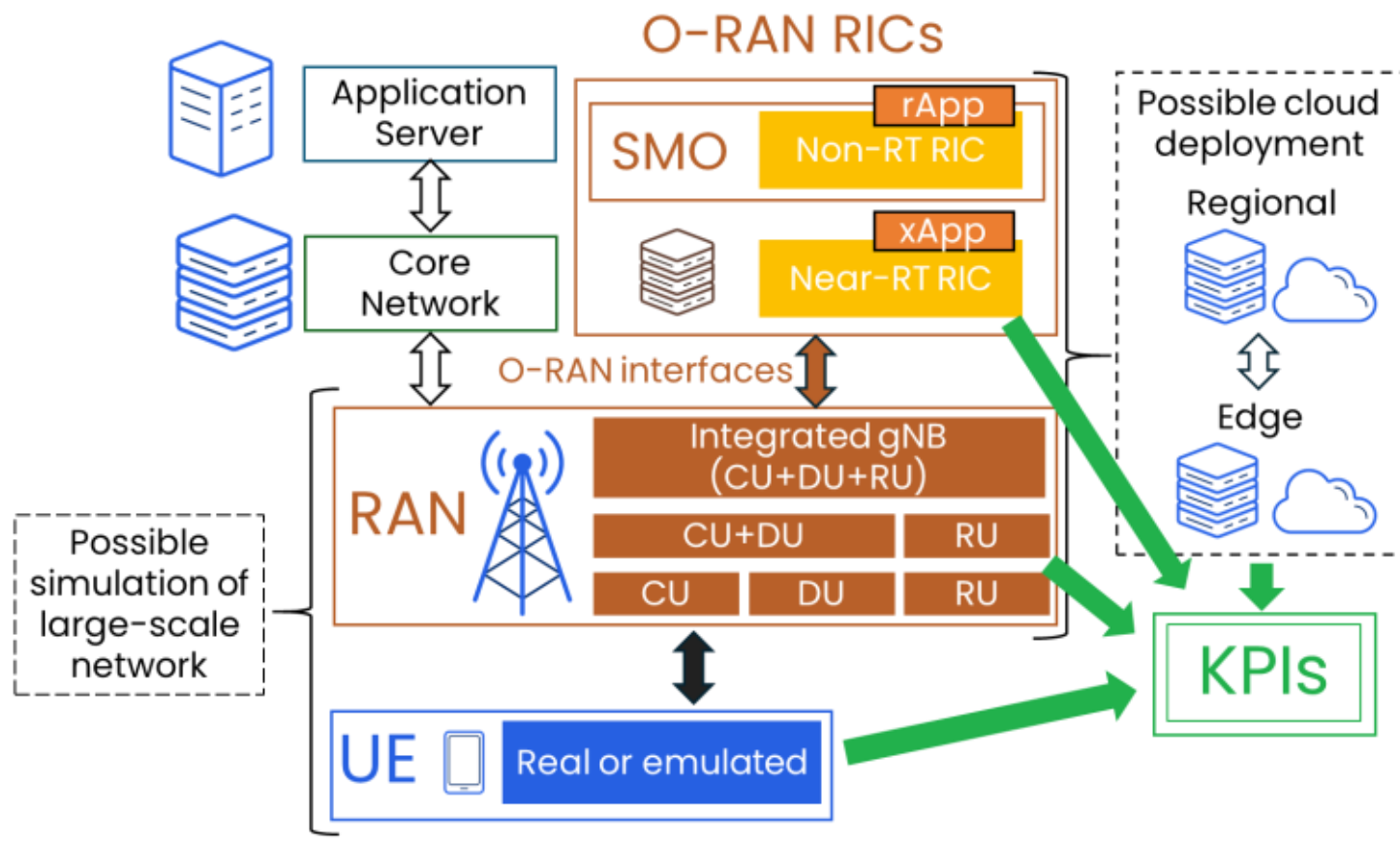

*Figure 1. E2E EE Testing Framework*

It can be seen that the O-RAN network enables a variety of deployment options for gNB, with combinations of O-RU/O-DU/O-CU, including their placement in the local or regional cloud. On top, there could be RICs with x/rApps. Moreover, in some cases, it would be beneficial to emulate the network using SW- or HW-based emulators of individual components, e.g., UE, O-RU, O-DU, or O-CU. There are multiple levels at which EE can be measured (e.g., RAN component, gNB stack, network-wide) and multiple devices under test (DUT), e.g., O-RU, O-DU, O-CU, RIC, x/rApp. Depending on the DUT and measurement scope, different components of the E2E EE Testing Framework can be selected and put together for different purposes. For example, one setup can be used to evaluate the EC/PC of an individual RAN component, like O-RU, while other aims at enabling performance comparison utilizing the RIC along with accompanying x/rApps. Taking this into account, below, we present four views on the E2E EE testing framework to capture various aspects.

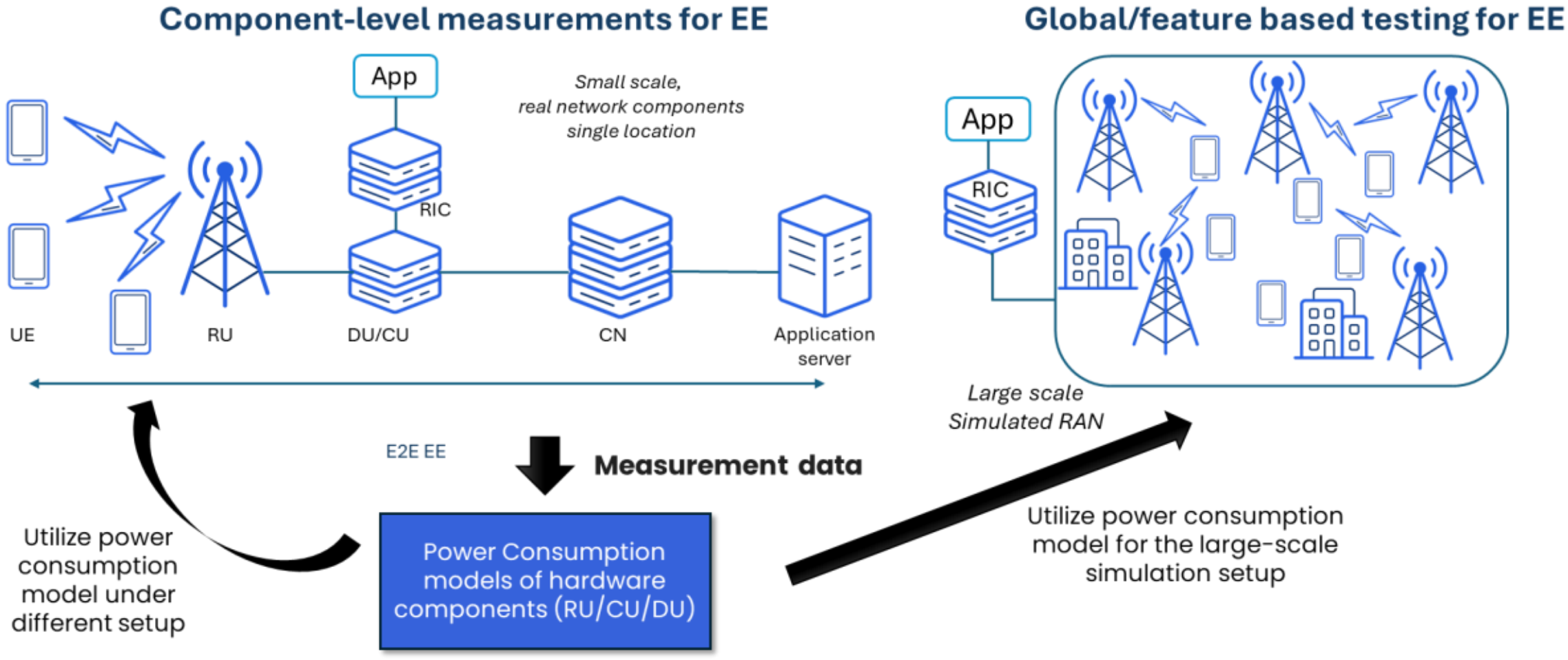

*Figure 2 Component-level and Global/feature-based views on E2E testing of EE, along with information exchange between them*

*A. Component Level View*

The component-level approach is intended for the evaluation of the EE of a single stack of the O-RAN components (see the left part of Fig. 2), i.e., O-RU, O-DU, or O-CU, inspired by the ETSI setup for BS PC measurements. We expect that this deployment utilizes the separately mentioned entities connected to the core network (possibly emulated). The UEs might be real or emulated. This setup can be used for the evaluation of the EE/EC of a single O-RAN component in a default configuration; xApps/rApps/RICs against the baseline approach.

*B. Global/Feature-Based View*

Global/feature-based option is intended for the evaluation of the EE algorithms deployed as x/rApps under a large, emulated network, e.g., hundreds of cells, under different scenarios and traffic profiles, e.g., rural, dense urban (see the right part of Fig. 2). For this kind of E2E EE evaluation, the unified models and definitions of scenarios are crucial. Moreover, the results of component-level measurements can be used here (e.g., PC models of O-RAN components provided by particular vendors), as well as the traffic models based on the MNOs' data. This setup can be used for evaluation of: the possible impact of conflicts; advanced x/rApps against a baseline set of x/rApps; x/rApps against baseline algorithms (i.e., algorithms within the traditional RAN).

*C. Deployment View*

Deployment view extends the component-level one with various O-RAN components deployed in different points of presence in a virtualized environment. It is intended to cover the EE aspects related to the cloud deployment options. The evaluation cases are the same as for the component-level view. However, measurement techniques are different, mostly relying on specific software. As an example, Redfish API exposed by the enterprise-grade servers can be used to track the PC of their HW components like fan, memory, or storage. In addition, the PC of CPU related to individual VNFs (processes, containers, or Kubernetes PODs) can be tracked using Kepler. The variety of possible variations of PNF and VNF, together with putting VNFs in the edge or regional cloud, makes the number of possible evaluation scenarios large.

*D. Multi-Link/Wide Network View*

Finally, multi-link/wide network view gathers all the above towards large-scale trials and serves as a big EE evaluation environment that utilizes multiple BSs together with RIC, and realistic cloud deployments. This could result in a large test network or a part of the commercial network. This option allows for conducting large-scale testing of x/rApps' influence on EE, as in the feature-based view, but in a real environment. Under such conditions, new aspects might impact the results, e.g., the delay between data reports and their processing by x/rApps deployed in the RIC, placed in the centralized cloud.

*E. Dependencies Between Views on EE Testing*

For practical reasons of creating a unified testing framework, we see the first two setups serving as a starting point for further work. Fig. 2 presents the relation and information exchange between the component-level and global views. While performing component-level EE testing, measurement data can be collected to formulate a parametrized PC model of a RAN component, like O-RU, O-DU/O-CU. This model can be transferred to the simulation environment utilized for the global/feature-based setup of EE testing to parameterize this component. Moreover, the obtained model can also be used again for component-level measurements, but in a different context, e.g., one can measure O-RU with an emulated O-DU and obtain a PC definition. Then the O-DU can be measured to obtain its PC model under an emulated O-RU, and use both models to combine the results to obtain E2E EE.

*F. Practical considerations*

One of the key aspects of the proposed framework is its practical applicability to cross-vendor benchmarking and vendor independence, with remarks below:

- the generic and modular definition of the framework allows measuring the EC at different entry points (e.g., O-CU, O-DU, O-RU, RIC, servers), allowing it to create various benchmarks to measure different aspects of EC;
- for measuring the specific aspect, e.g., comparing the performance of ES Apps, the setup shall be fixed – i.e., a benchmark, using e.g., a single RAN, or RAN emulator and RIC - when the aim is to compare the performance of the RAN under a single setup of RIC and Apps. Another one could be to measure the behavior and performance of different RICs, while the Apps and RAN should be fixed.

Moreover, practical deployment encounters the following challenges:

- latency and signaling overhead;
- real-time data collection constraints;
- changing configuration of the network and dynamic traffic characteristics – with small and large traffic volumes, which imply intelligent averaging of the measurements;
- measurement accuracy;
- vendor-specific limitations and variations of the response time, etc.

The above needs to be taken into account when extrapolating the potential EE gains in the real networks. To this end, the Multi-link/Wide-Network View (presented in Sec. III D.) was proposed, which reflects the close-to-real network and larger-scale effects. The bigger the setup, the closer to a real network the measurement results. This, however, significantly increases the complexity and cost of the framework application, yielding this to be a conceptual approach. From this perspective, the smaller setups can be more financially affordable in practice, which, on the contrary, might limit the applicability of the results in laboratory setups towards the expected energy consumption in real network.

## IV. Example of Global/Feature-Based View on E2E EE Testing in O-RAN

To verify the applicability of the proposed framework presented in the previous section, we utilize it for x/rApp EE testing under the large-scale simulated network. For this purpose, we followed the Global/Feature-Based view, which is a straightforward setup in a lab environment, e.g., compared to the Deployment or Multi-Link/Wide Network view.

*A.EE Testing Setup*

For the practical realization of Global/Feature-based testing of EE, we utilized a carrier-grade RAN emulator provided by Keysight Technologies, i.e., RICTest, simulating the whole O-RAN protocol stack in real-time, including E2, O1, and 3GPP-related interfaces. We deployed 39 cells based on the real-world data provided by Deutsche Telekom, with twelve three-sector sites and three one-sector sites. The test duration was 60 minutes, supporting a varying number of UEs with a steady throughput demand of 4 Mbps in the uplink and 40 Mbps in the downlink. During the test timeline, there were 3 states, each lasting 20 minutes, defined as follows:

- **Low load state** with 160 UEs;
- **Transition state** with the UE number growing from 160 to 390 in four steps;
- **High load state** with 390 UEs.

Such a configuration allows for evaluation of the EE improvement resulting from the use of xApps and rApps under low load conditions, as well as their ability to adapt to the growing network load. As a first resultant KPI, we utilized EE as per the definition in [3]:

$$EE = \frac{total\ data\ volume}{energy\ consumption}\left[\frac{bit}{J}\right].$$

EE was measured over time per whole network. In addition, to investigate the impact of EE optimization on the UE performance, we introduced the QoS score KPI:

$$QoS\ score = \frac{average\ UE\ throughput}{UE\ throughput\ demand}\,[\%].$$

The network EE and QoS score achieved in a setup with x/rApps under test was compared against the baseline scenario without x/rApps deployed, i.e., all cells were active, and no energy-saving algorithms were deployed in the network. The test setup was hosted in the i14y Lab in Germany, and its parameters are summarized in Table 2.

*Table 2 Parameters of EE testing setup (based on [15])*

| Org. | Scope |
|---|---|
| **RAN Emulator** | RICTest by Keysight (v4.3) |
| **O-RAN Interfaces** | O1, E2, A1, R1 |
| **Cell Deployment** | • 12 sites with 3 sectors<br>• 3 sites with 1 sector<br>• 39 cells in total<br>• 64TxRx mMIMO used for each cell |
| **Spectrum Setup** | • 5G NR n77 band (TDD)<br>• 100 MHz bandwidth |
| **UE Deployment** | • 390 UEs in total<br>• 4 Mbps throughput demand in UL<br>• 40 Mbps throughput demand in DL |
| **Test Duration** | 60 minutes |

| Traffic Profile | • **Low load state** (20 minutes, 160 UEs, 10% of maximal system capacity)<br>• **Transition state** (20 minutes, UEs growing from 160 to 390 in four steps)<br>• **High load state** (20 minutes, 390 UEs, 30% of maximal system capacity) |
|---|---|
| KPIs | • **EE**<br>• **QoS score** |

*A. Evaluation of the Tandem of Cell On/Off Switching xApp and rApp*

Using the proposed E2E EE Testing Framework, and following the Global/Feature-based view described above, it is possible to evaluate EE of xApps and rApps under use-cases defined by O-RAN ALLIANCE [10]:

- **Carrier and Cell Switch On/Off Switching:** switching on/off entire cell or component carrier to provide ES.
- **RF Channel Reconfiguration:** dynamically switching on/off some TxRx chains in the mMIMO system for ES.
- **Advanced Sleep Mode Selection:** providing ES by switching on/off certain parts of the base station, like the power amplifier, in very short cycles of symbol slot or frame.

This is enabled by the varying traffic load forcing xApps/rApps under test to take advantage of low load for improving EE, and during high load, fulfill UE QoS demand. While the cells are equipped with mMIMO antenna arrays, it is not only possible to test carrier and cell on/off switching, but also RF channel reconfiguration, or advanced sleeping modes. For each use case, the EE gains and QoS score can be evaluated by comparing against the baseline scenario with no EE optimization.

As a representative example of how the proposed E2E EE Testing Framework can be used, we tested a hierarchical approach to the cell on/off switching using a tandem of xApp-rApp developed by Rimedo Labs. The xApp decides locally on turning certain cells on/off, while being guided by the rApp through the A1 policies formulated based on the wide-network KPIs. The A1 policies contain a set of thresholds to either encourage xApp to switch off cells to save energy, or to switch them back on to maintain user QoS [15]. The xApp and rApp are deployed in the Near-RT and Non-RT RIC platform by Juniper Networks, respectively. The outcome of EE tests is depicted in Fig. 3. The EE of App-tandem under test is marked with a magenta line and is compared against the baseline (without EE optimization) marked with a grey line. The result of the test shows that compared to the baseline, the xApp-rApp tandem can significantly improve the EE during the low load state, and during both the transition and high load states, EE is not deteriorated. In detail, during the low load state, up to 57% improvement of EE can be achieved compared to the baseline. In high load state, the EE of the tested xApp-rApp tandem and baseline is almost the same. However, this is expected behavior, as the network must ensure the required QoS for a growing number of users. A deeper look into the QoS is presented in Fig. 4. In essence, the throughput of users is not affected by the switching off of cells during the low load state. Although at the beginning of the transition state, a degradation of the QoS score to about 75% is observed, the App-tandem reacts properly and improves the QoS score by switching on cells to reach and maintain the value of about 96% during the high load state. Overall, the drop in QoS score to 75% persisted for about 4 minutes. It is quite a significant value; thus, the MNO would most likely set the xApp-rApp operation to act more "lightly" to minimize the QoS drop, as this may be too high cost of an EE gain to be acceptable. This test can be treated as a representative example of how the proposed E2E EE Testing Framework could be used to evaluate the xApps and rApps in the large-scale network. The key parts of this setup are that it is based on the real-world data from the MNO, that it utilizes the commercially available O-RAN compliant RICs, and that it utilizes a carrier-grade RAN emulator, ensuring compliance with O-RAN protocols. Most importantly, the presented benchmarking example is one of the possible realizations of the general E2E EE Testing Framework. Depending on the available SW, the test can be repeated, e.g., using the same xApps and rApps and a different RIC platform, or different xApps/rApps under the same RIC platform. Moreover, the same traffic/cell deployment scenario can be reproduced under different RAN emulators. Finally, more evaluation scenarios can be defined based on additional data provided by MNO while keeping the Apps, RIC, and RAN emulator.

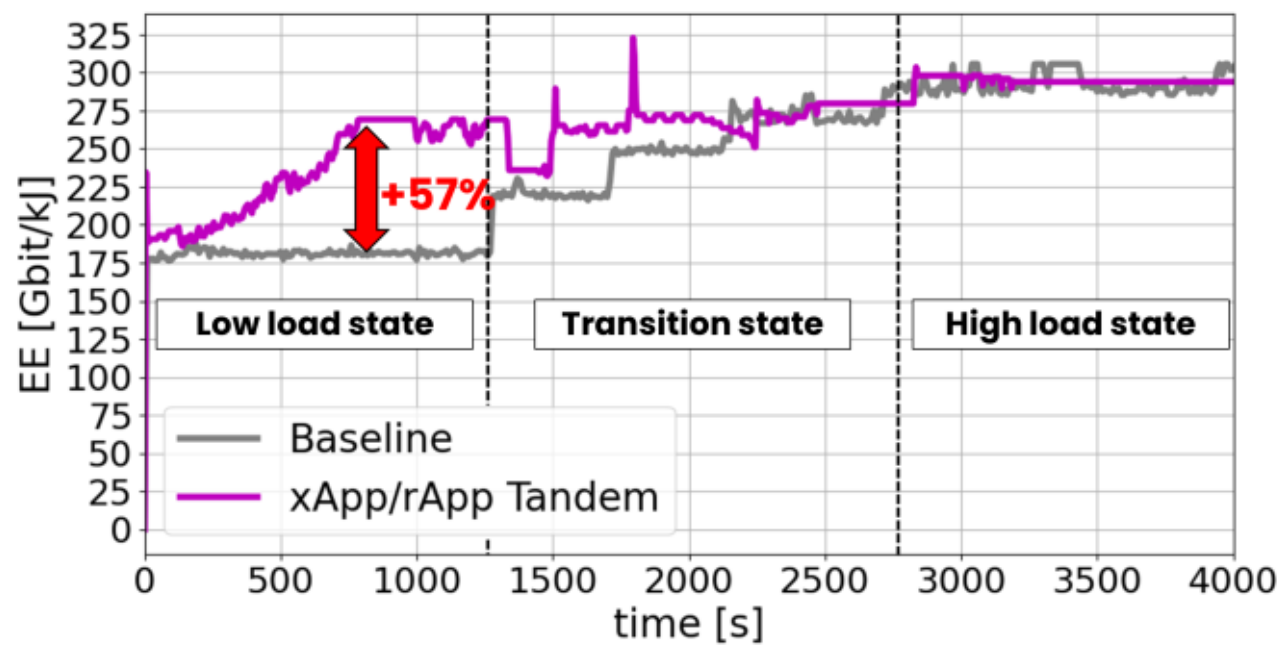

*Figure 3 EE of tandem of xApp and rApp obtained using the proposed E2E EE Testing Framework*

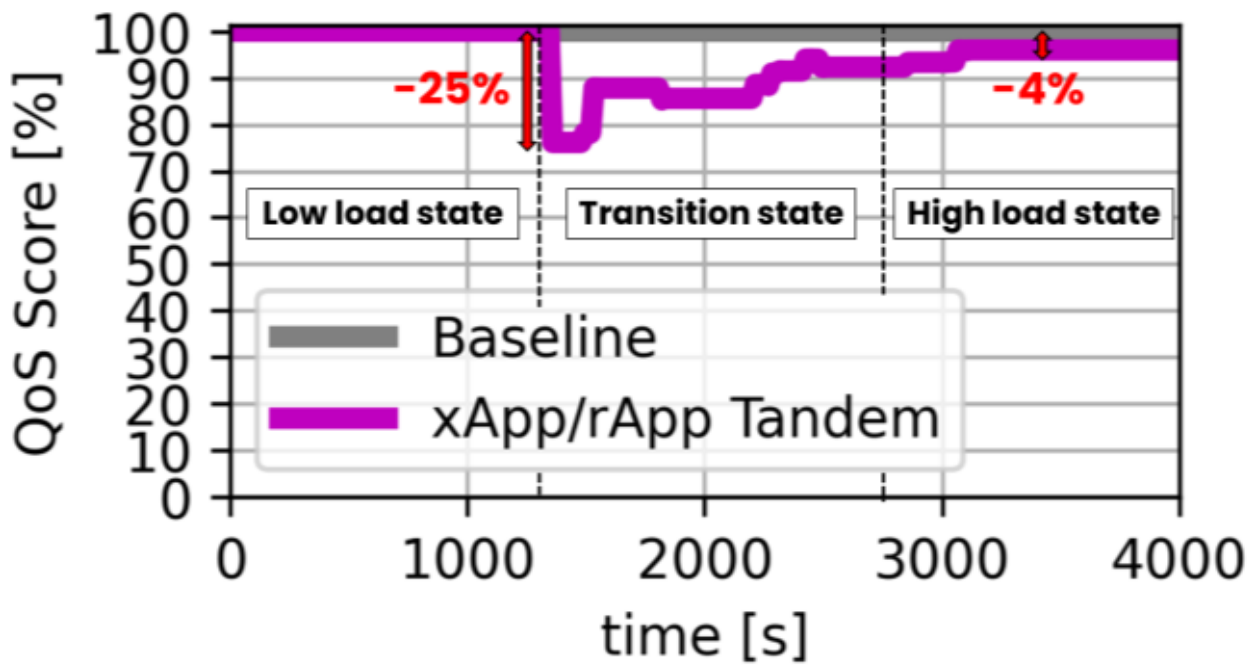

*Figure 4 QoS score of the tandem of xApp and rApp obtained using the proposed E2E EE Testing Framework*

## V. Conclusions

The paper proposes an E2E EE testing framework for mobile networks, highlighting key aspects and challenges. As the analyzed documents from various standardization-related organizations address separate aspects of EE, the authors propose to combine the outcomes into a unified framework. The paper identifies four views of EE testing, namely: a realistic small-scale setup with interchangeable components to test the EE performance of individual elements; large-scale simulation or emulation to evaluate the performance of energy-saving x/rApps; a setup with cloud deployment options, such as edge or regional sites, to assess EE under diverse network elements locations; and a large-scale realistic setup with different cloud deployments to conduct comprehensive EE testing. As proof of concept, the practical realization of the proposed E2E EE Testing Framework is presented, covering the Global/Feature-based view on EE testing. The setup utilizes real-world data from an MNO, commercial RICs, and a RAN emulator. It is then used to evaluate the EE of xApp/rApp Tandem by Rimedo Labs, showing the gains of up to 57% during the low load state. Future work for the framework's design and execution should involve developing a standardized testing procedure adaptable to various scenarios, ensuring reliability and comparability of the results, and avoiding excessive configurability of modules to prevent inconsistencies.

## Acknowledgment

This work was sponsored and supported by the i14y Lab - a multi-party research and development project with funding by the German Federal Ministry for Digital and Transport (BMDV).

## Literature

[1] M. Hoffmann, M. Dryjański, "Energy Efficiency in Open RAN: RF Channel Reconfiguration Use Case," IEEE Access, vol. 12, 2024

[2] O-RAN ALLIANCE, O-RAN Global Plugfest Fall 2024, available online: https://plugfestvirtualshowcase.o-ran.org/2024/Joint_European_O-RAN_PlugFest accessed: 26.06.2025

[3] ETSI TS 103 786, "Measurement method for energy efficiency of wireless access network equipment; Dynamic energy efficiency measurement method of 5G Base Station (BS)", v1.2.1

[4] ETSI ES 202 706-1, "Metrics and measurement method for energy efficiency of wireless access network equipment; Part 1: Power consumption - static measurement method", v1.7.1

[5] ETSI ES 203 539, "Environmental Engineering (EE); Measurement method for energy efficiency of Network Functions Virtualisation (NFV) in laboratory environment", v1.1.1

[6] 3GPP TR 38.864, "Study on network energy savings for NR", v18.1.0

[7] 3GPP TR 26.942, "Study on Media Energy Consumption Exposure and Evaluation Framework", v0.3.0

[8] 3GPP TR 32.927, "Study on system and functional aspects of energy efficiency in 5G networks", v18.0.0

[9] 3GPP TR 38.901, "Study on channel model for frequencies from 0.5 to 100 GHz", v18.0.0

[10] O-RAN ALLIANCE, WG1, "Network Energy Saving Use Cases Technical Report", v02.00

[11] O-RAN ALLIANCE, TIFG, "O-RAN End-to-End System Testing Framework Specification 1.0", v01.00

[12] O-RAN ALLIANCE, WG6, "Study on O-Cloud Energy Savings", v01.00

[13] O-RAN ALLIANCE, WG7, "Network Energy Savings Procedures, and Performance Metrics", v01.00

[14] NGMN Alliance, "Green Future Networks: Metering in Virtualized RAN Infrastructure", v1.0, April 2024

[15] Rimedo Labs, T-Labs, "Multi-scale hierarchical rApp-xApp tandem for Energy Saving using real mobile network data" O-RAN Virtual Exhibition MWC Barcelona 2025, available online: https://shorturl.at/ITkid, accessed: 26.06.2025

**Marcin Hoffmann** is a Technical Solution Manager at Rimedo Labs, working on O-RAN software development solutions. He is also a Ph.D. candidate at PUT. His research includes utilizing machine learning for 5G/6G network management.

**Marcin Dryjański** serves as CEO and principal consultant at Rimedo Labs. He received his Ph.D. from the Poznan University of Technology. He is a co-author of many articles on 5G and Open

RAN, and a co-author of the book „From LTE to LTE-Advanced Pro and 5G" (M. Rahnema, M. Dryjanski, Artech House 2017).

**Adrian Kliks** is an assistant professor at the Institute of Radiocommunications. His main fields of interest are Open RAN implementations and radio resource management.

**Andreas Gladisch owns** PhD in Optical Communication from Humboldt University Berlin. He has a long track record inside Deutsche Telekom as Vice President in the field of Telecom Innovation and Technology and is the consortium lead of the i14y Lab project.

**Ajesh Pulyaar Keerthi** works for Deutsche Telekom Group Technology in the i14y Lab as a Network Architect and 5G Test engineer.

**Mohammadreza Razmi** works for Deutsche Telekom Group Technology in the i14y Lab as a Network Architect and 5G Test engineer.

**Heiko Lehmann** received the Diploma degree in physics in 1988 and the Ph.D. degree in theoretical physics in 1992, both from Humboldt University of Berlin, Berlin, Germany, respectively. Currently, he works as Tribe Lead Digital Twin & Cybersecurity Group Technology in Deutsche Telekom T-Labs.